# Dependence of the gap parameter on the number of CuO$_2$ layers in a unit cell of optimally doped BSCCO, TBCCO, HBCCO and HSCCO

Ya G Ponomarev[1], N Z Timergaleev[1], A O Zabezhaylov[1], Kim Ki Uk[1], M A Lorenz[2], G Müller[2], H Piel[2], H Schmidt[2], C Janowitz[3], A Krapf[3], R Manzke[3]

[1] M.V. Lomonosov Moscow State University, Faculty of Physics, 119899 Moscow, Russia
[2] Bergische Universität Wuppertal, Fachbereich Physik, Gaußstr. 20, D-42097 Wuppertal, Germany
[3] Humboldt-Universität zu Berlin, Institut für Physik, Invalidenstr. 110, D-10115 Berlin, Germany

**ABSTRACT:** We have measured the superconducting gap $\Delta_s$ in optimally doped samples of $Bi_2Sr_2Ca_{n-1}Cu_nO_{2n+4}$, $Tl_2Ba_2Ca_{n-1}Cu_nO_{2n+4}$, $HgBa_2Ca_{n-1}Cu_nO_{2n+2}$ and $HgSr_2Ca_2Cu_3O_8$ by Andreev and tunnelling spectroscopy and by ARPES. We have found that the low-temperature value of $\Delta_s$ within experimental errors is linearly increasing with the number $n$ of CuO$_2$ layers in the unit cell of the investigated HTSC-families ($n \leq 3$). The variation of the critical temperature $T_{c\,max}$ with $n$ does not obey this simple relation.

## 1 INTRODUCTION

It is generally assumed that in the layered high-$T_c$ superconductors (HTSC) $Bi_2Sr_2Ca_{n-1}Cu_nO_{2n+4+\delta}$ (BSCCO), $Tl_2Ba_2Ca_{n-1}Cu_nO_{2n+4+\delta}$ (TBCCO) and $HgBa_2Ca_{n-1}Cu_nO_{2n+2+\delta}$ (HBCCO) the superconductivity in a unit cell is located in blocks of $n$ closely spaced CuO$_2$ layers intercalated by Ca (Waldram 1996). These superconducting blocks are separated in **c**-direction by insulating or semiconducting blocks (spacers) with a structure universal for a given HTSC-family. The spacers play an important role in the formation of superconducting properties of the above mentioned compounds serving as charge reservoirs. For a phase with a given number of CuO$_2$ layers $n$ the maximal critical temperature $T_{c\,max}$ (and correspondingly the maximum superconducting gap $\Delta_{s\,max}$ (Deutscher 1999 and Ponomarev 1999b)) can be achieved by the variation of the excess oxygen concentration $\delta$ in the spacers. The excess oxygen attracts electrons from the CuO$_2$ layers without introducing a significant scattering of charge carriers in the superconducting blocks ("distant" doping). In addition the excess oxygen forms traps in the spacers, thus strongly influencing the transport in **c**-direction due to resonant tunnelling effects (Abrikosov 1999 and Halbritter 1996).

|  |  | $T_c$ [K] | $\Delta_s$(4.2 K) [meV] | $2\Delta_s/kT_c$ | Reference |
|---|---|---|---|---|---|
| BSCCO | $n=1$ | 25 ± 2 | 12.7 ± 0.5 | 11.8 ± 1.1 | this work |
|  | $n=2$ | 86 ± 4 | 25 ± 1 | 6.7 ± 0.5 | this work |
|  | $n=3$ | 110 ± 5 | 36 ± 1.6 | 7.6 ± 0.5 | this work |
| TBCCO | $n=1$ | 73 ± 2 | 14 ± 1 | 4.5 ± 0.4 | Tsai et al (1989) |
|  | $n=2$ | 104 ± 5 | 34 ± 1 | 7.6 ± 0.5 | this work |
|  | $n=3$ | 118 ± 5 | 47.7 ± 1.4 | 9.4 ± 0.5 | this work |
| HBCCO | $n=1$ | 94 ± 2 | 31 ± 1 | 7.7 ± 0.5 | this work |
|  | $n=2$ | 124 ± 5 | 49 ± 1.5 | 9.2 ± 0.5 | this work |
|  | $n=3$ | 135 ± 5 | 75 ± 1.5 | 13 ± 0.6 | Wei et al (1998) |
| HSCCO | $n=3$ | 107 ± 5 | 36 ± 1.5 | 7.8 ± 0.5 | this work |

Table 1. Superconducting properties of the investigated HTSC compounds.

Samples with the maximum critical temperature $T_{c\,max}$ corresponding to the excess oxygen concentration $\delta = \delta_{opt}$ are defined as optimally doped. For the optimally doped HTSC samples the dependence of the critical temperature $T_{c\,max}$ on the number of $CuO_2$ layers $n$ is strongly nonlinear (Phillips 1994). Several theoretical models have been proposed to explain the nontrivial dependencies $T_{c\,max}(n)$ in HTSC materials (Phillips 1994, Byczuk et al 1996, Kresin et al 1996, Chen et al 1997 and Leggett 1999). Unfortunately this problem remains unsolved mainly because of the limited and often conflicting experimental data on the variation of superconducting properties of HTSC with the number of $CuO_2$ planes in the superconducting blocks (Hudáková et al 1996 and Wie 1998).

This paper presents low-temperature measurements of the superconducting gap $\Delta_s$ as a function of $n$ ($n \leq 3$) in optimally doped samples of the BSCCO-, TBCCO- and HBCCO-families with identically constructed superconducting blocks. A correlation of the obtained $\Delta_s(n)$-dependencies with the structure of spacers in these families has been found.

## 2 EXPERIMENTAL RESULTS AND DISCUSSION

We have measured the superconducting gap $\Delta_s$ in optimally doped samples of BSCCO ($n = 1, 2, 3$), TBCCO ($n = 2, 3$), HBCCO ($n = 1, 2$) and $HgSr_2Ca_2Cu_3O_{8+\delta}$ (HSCCO, $n = 3$) by Andreev and tunnelling spectroscopy at $T = 4.2$ K with the current in **c**-direction using a break junction technique (Aminov et al 1996). The optimally doped samples of BSCCO ($n = 2, 3$) have been studied also by angle-resolved photoemission spectroscopy (ARPES) at $T \approx 30$ K. The details of ARPES measurements have been published elsewhere (Müller et al 1997).

Point-contact and tunnelling spectroscopy studies of the optimally doped single crystals of $Bi_2Sr_{2-x}La_xCu_1O_{6+\delta}$ gave exactly the same value of the gap $\Delta_s(4.2\,K) = (12.7 \pm 0.5)$ meV (Ponomarev et al 1999b) (Table 1), which is in reasonable agreement with the results of the earlier ARPES measurements of $\Delta_{s\,max}(9\,K) = (10 \pm 2)$ meV by Harris et al (1997). Similar measurements have been performed on the optimally doped single crystals of BSCCO ($n = 2$) and polycrystalline samples of BSCCO ($n = 3$). In the $dI/dV$-characteristics of SNS contacts (point-contact regime) a clearly defined subharmonic gap structure (SGS) caused by multiple Andreev reflections has been observed for all investigated samples (Fig. 1, Fig. 2). In all cases, the dips at bias voltages $V_n = 2\Delta/en$ composing the SGS had a symmetric form which points to the absence of a strong anisotropy of the gap $\Delta_s$ in the **ab**-plane (Devereaux et al 1993). The value $\Delta_s(4.2\,K) = (25 \pm 1)$ meV obtained in this work for BSCCO ($n = 2$) from Andreev and tunnelling spectroscopy (Table 1) is well supported by recent studies of intrinsic Josephson effect ($\Delta_s = 25$ meV) (Ponomarev et al 1999a and Suzuki et al 1999) and is in agreement with the ARPES measurements by Müller et al (1997) ($\Delta_{s\,max}(30\,K) = (28 \pm 2)$ meV).

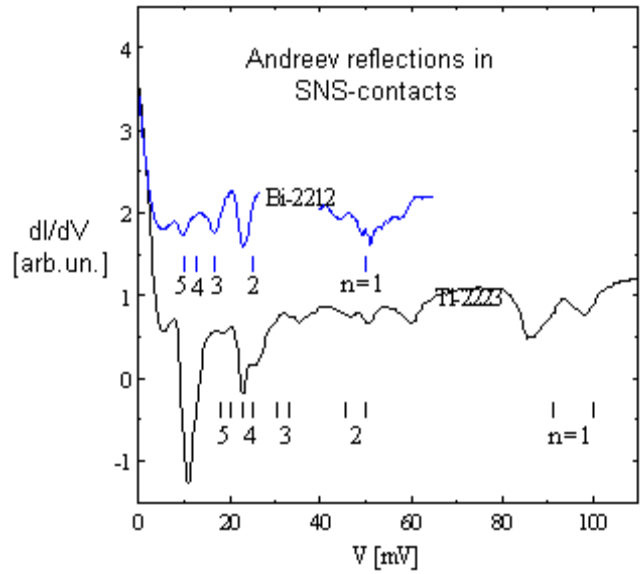

Fig.1. Subharmonic gap structure (SGS) in the d$I$/d$V$-characteristics of SNS contacts in an optimally doped single crystals of BSCCO ($n$ = 2) and TBCCO ($n$ =3) at $T$ = 4.2 K. The bias voltages $V_n$ = 2$\Delta_s$/$en$ corresponding to the dips in SGS are marked by bars.

We have also estimated the values of $\Delta_s$ for the TBCCO ($n$ = 2, 3) single crystals from Andreev spectroscopy measurements at $T$ = 4.2 K. The same was done for the optimally doped polycrystalline samples of HBCCO ($n$ = 1,2). All the resulting values are summarised in Table 1. It should be noted that the experimental data obtained in this work from Andreev and tunnelling spectroscopy corresponded to the current in the **c**-direction in case of single crystals.

Recently an impressive experimental evidence supported by sound theoretical considerations has been presented which showed that in layered HTSC crystals a s-wave component of the order parameter is dominating for the transport in the **c**-direction (Klemm et al 1999). Probably the d-wave component is filtered out due to the specific character of this transport. This view is in accordance with recent studies of the intrinsic Josephson effect in BSCCO (Ponomarev et al 1999a). Furthermore it has been pointed out by Devereaux and Fulde (1993) that in the case of an anisotropic s-wave order parameter the SGS in the d$I$/d$V$-characteristic of a SNS junction should split into two separate structures corresponding to $V_{n1}$ = 2$\Delta_{min}$/$en$ and $V_{n2}$ = 2$\Delta_{max}$/$en$. Actually we have often observed such splitting for the best SNS contacts in all three investigated HTSC-families (see for example the SGS for the Tl-2223 contact in Fig 1). If the version with the anisotropic s-wave order parameter is correct then the ratio $\Delta_{max}$/$\Delta_{min}$ in the samples of the investigated HTSC-families does not exceed ($\Delta_{max}$/$\Delta_{min}$) = 1.1 at $T$ = 4.2 K.

Combining our results with the data obtained earlier for HBCCO ($n$ = 3) (Wei et al,1998) and TBCCO ($n$ = 1) (Tsai et al,1989), we have plotted the gap $\Delta_s$ vs $n$ for all three HTSC-families (Fig. 3). Obviously the gap $\Delta_s$ in the investigated HTSC-families is proportional to the number of CuO$_2$ layers $n$ in the range $n \leq 3$. It should be noted that the critical temperature $T_{c\,max}$ vs $n$ does not obey this simple relation (Phillips 1994).

In conclusion we have found that the low-temperature value of a superconducting gap $\Delta_s$ is linearly increasing with the number $n$ of CuO$_2$ layers in the unit cell of optimally doped samples of the BSCCO-, HBCCO- and TBCCO-families ($n \leq 3$)(Fig.3). The variation of the critical temperature $T_{c\,max}$ with $n$ does not obey this simple relation.

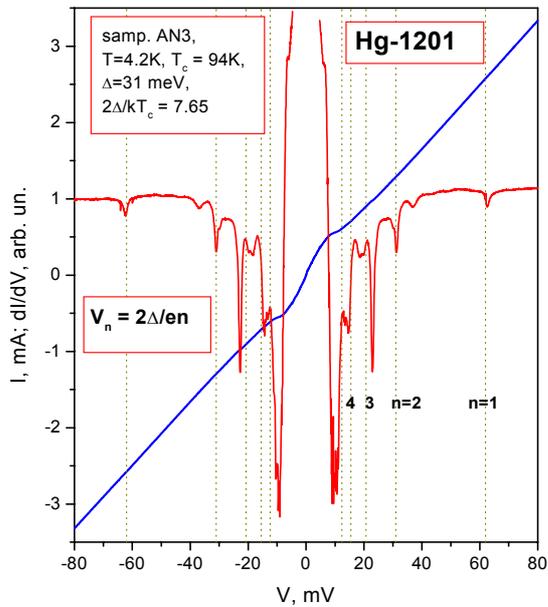 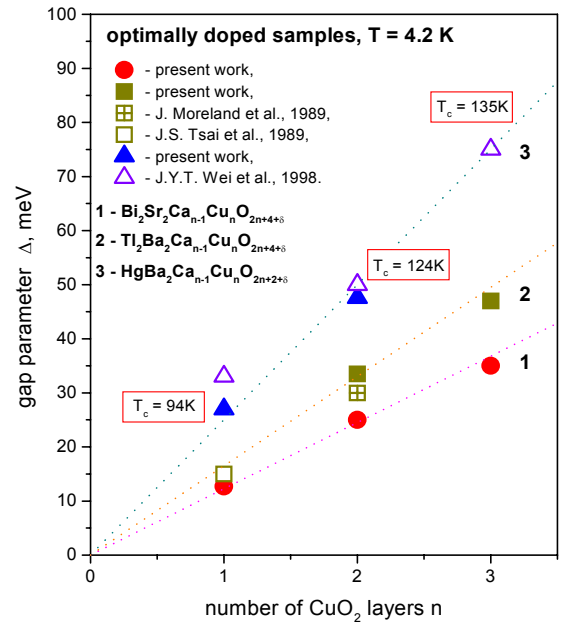

Fig.2. Subharmonic gap structure (SGS) in the d$I$/d$V$-characteristics of SNS contacts in an optimally doped polycrystalline sample of HBCCO ($n = 2$) at $T = 4.2$ K. The bias voltages $V_n = 2\Delta_s/en$ corresponding to the dips in SGS are marked by bars.

Fig.3. The superconducting gap $\Delta_s$(4.2 K) vs the number of CuO$_2$ layers $n$ for the optimally doped samples of BSCCO-, TBCCO- and HBCCO-families.

**ACKNOWLEDGMENTS**

The authors would like to thank K. Winzer for supplying high-quality TBCCO ($n = 3$) single crystals. The authors are indebted to T.E. Os'kina, Yu.D. Tretyakov and N. Kiryakov for providing the BSCCO ($n = 2$) crystals and the polycrystalline samples of BSCCO ($n = 3$) and HBCCO ($n = 2, 3$). This work was supported in part by the ISC on High Temperature Superconductivity (Russia) under the contract number 96118 (project DELTA) and by RFBR (Russia) under the contract number 96-02-18170a.